\documentclass[journal]{IEEEtran}
\usepackage{graphicx}
\usepackage{dcolumn}
\usepackage{bm}
\usepackage{caption}
\usepackage{comment}
\usepackage{subcaption}
\usepackage{xcolor}
\usepackage{physics}
\usepackage{ulem}
\usepackage{placeins,citesort}
\usepackage{amsmath,amssymb}

 \usepackage{algpseudocode}

\usepackage{algorithm}

 \newcommand{\dbar}[1]{\bar{\bar{#1}}}
\begin{document}

\title{Domain Decomposition Framework for Maxwell Finite Element Solvers and Application to PIC}

\author{Zane~D.~Crawford,~\IEEEmembership{Student~Member,~IEEE,}
        O.~H.~Ramachandran,~\IEEEmembership{Student~Member,~IEEE,}
        Scott~O'Connor,~\IEEEmembership{Student~Member,~IEEE,}
        Daniel L. Dault,~\IEEEmembership{Member,~IEEE,}
        John~Luginsland,~\IEEEmembership{Fellow,~IEEE,}
        and~B.~Shanker,~\IEEEmembership{Fellow,~IEEE}
        
        \thanks{  Z. D. Crawford, O. H. Ramachandran, S. O'Connor, B. Shanker are with the Department
of Electrical and Computer Engineering, Michigan State University, East Lansing,
MI, 48824.\protect\\
D. Dault is with AFRL, Dayton, OH 4543\protect\\
 J. Luginsland is with AFRL/Air Force Office of Scientific Research, Arlington, VA 22201.
E-mail: bshanker@egr.msu.edu}
}
\markboth{IEEE Transactions on Plasma Sciences,~Vol.~14, No.~8, Feb~2020}%
{Shell \MakeLowercase{\textit{et al.}}: Bare Demo of IEEEtran.cls for IEEE Journals}

\maketitle
\begin{abstract}
The most popular methods for self-consistent simulation of fields interacting with charged species is using finite difference time domain (FDTD) methods together with Newton's laws of motion to evolve locations and velocities of particles. Despite their popularity, the limitation of FDTD particle in cell (EM-FDTDPIC) methods are well known. To address these, there has been significant interest over the past decade in exploring alternatives. In the past few years, the advances in electromagnetic finite element methods for particle in cell (EM-FEMPIC) has advanced by leaps and bounds. The mathematics necessary for implicit FEM methods that are unconditionally stable and charge conserving are now well understood. Some of these advances are more recent. The next bottleneck necessary to make EM-FEMPIC competitive with FDTD based scheme is overcoming computational cost. Our approach to resolving this challenge is develop two different finite element tearing and integration approaches, and using these to create domain decomposition schemes for EM-FEMPIC. Details of the proposed methodology are presented as well as a number of results that demonstrates charge conservation as well as amelioration of costs for a number of problems. 
\end{abstract}


\IEEEpeerreviewmaketitle

\section{Introduction}\label{sec:introduction}

The design of novel accelerator and vacuum devices require computational tools capable of simulating self-consistent interactions between plasma and electromagnetic fields \cite{shin2010particle,abrams2001vacuum,martins2010exploring,dolgashev2001simulations}.
This is commonly done by evolving a particle distribution in time due to the Lorentz force defined by an electric field and magnetic flux density; this includes both external/impressed and those generated by the acceleration of charged species. The predominant approach is the finite difference time domain based particle in cell method (EM-FDTDPIC) \cite{birdsall2018plasma}. This method has seen decades of use and advancement, primarily due its ease of use and efficiency, as the fields can be determined from a simple recurrence formula rather than matrix inversion. Furthermore, parallel implementations and fast techniques have enabled EM-FDTDPIC to simulate large problems \cite{sasser1999current,vay2012novel,fonseca2002osiris,qiang2000object}.
However, there has been interest in exploring alternatives to EM-FDTDPIC. To a large extent, this is motivated by the higher order fidelity that one can obtain in terms of both geometry as well as physics. 

In response to this need, the past decade has seen extensive effort to create an electromagnetic particle in cell method based around the finite element method (EM-FEMPIC); see  \cite{meierbachtol2015conformal,pinto2014charge} and papers therein. It has been shown that by choosing proper basis sets for the fields and currents, it is possible to obtain charge conserving EM-FEMPIC schemes  through exact current mapping \cite{moon2014exact,o2021set,pinto2014charge}. Unfortunately, these methods use explicit time stepping with first order basis functions to represent both the fields, paticles, currents, and time. More recent work has utilized higher order spatial basis better represent fields, but the error in representation is still constrained by the representation for time \cite{glasser2019geometric}; note, the moniker ``structure-preserving'' is used when referring to a method that uses Whitney basis/ensures that the \emph{de-Rham} relations hold. The downside of using explicit time stepping is that it is \emph{conditionally stable}, and the time step to ensure stability is determined by the smallest feature size. Obviously, this is a challenge when analyzing plasma in geometrically complex in structures. 

More recent work provided a general framework for charge conservation to be satisfied, irrespective of time stepping scheme \cite{crawford2021rubrics}. A practical demonstration of satisfaction of conservation laws for implicit unconditionally stable finite element scheme for Maxwell equation was shown for a set of test problems in \cite{o2021time}. It was also shown in this paper, that the proposed methodology was valid for a wave equation FEM solver as well. Unfortunately, it is well known that there exists a null space that grows as $t \grad \phi (\vb{r})$ for the  wave equation solvers and a null space that varies as $\grad \phi (\vb{r})$ for implicit Maxwell solvers, where $\phi (\vb{r})$ is a scalar field. While it may be possible to control the magnitude of the null space excited in Maxwell solvers, the one for the wave equation unfortunately rears its head with time. The development of a discrete Coulomb gauge in \cite{o2021quasi} shows how one may obtain a method that satisfies Gauss' laws for \emph{both} wave equation and Maxwell equation solvers. Despite this progress, a fundamental bottleneck remains. All EM-FEMPIC methods are more expensive than EM-FDTDPIC methods. 

The reason for the efficiency of FDTD schemes fairly simple; there is no inversion thanks to the structure of the method \cite{tonti2001finite}. Unfortunately, FEM methods typically involve inversion of matrices. As is well known, a naive direct inversion would cost $\mathcal{O}(N_{dof}^3)$ per time step where $N_{dof}$ is the number of degrees of freedom. This can be reduced to $\mathcal{O}(N_{dof} \log N_{dof})$ if a multifrontal sparse solver is used \cite{duff1983,butterMultifrontal} with $\mathcal{O}(N_{dof}^\nu polylog (N_{dof}) )$ for factorization where $\nu < 2$ \cite{multifrontal1}. Alternatively, one typically uses an iterative solver whose cost scales as $\mathcal{O}(N_{iter}N_{dof})$, where $N_{iter}$ is the number of iterations which depends on the error threshold. As a result, one takes recourse to preconditioners. But it has been shown that the most computationally effective approach has been to use domain decomposition \cite{farhat1991method,farhat2001feti}. A bulk of the effort has been applying this technique to amortize the cost of wave equations and largely in the Fourier domain \cite{li2006vector,li2008implementation,xue2012nonconformal,wolfe2000parallel}, with some scant papers in the time domain \cite{akbarzadeh2019unconditionally,navsariwala1997efficient}. The primary benefit of this approach is to reduce the complexity of the problem by effectively reducing the size of $N_{dof}$ that needs to be inverted.

The domain decomposition approach in this work is the finite element tearing and interconnecting (FETI) framework. Here, the domain is partitioned into multiple regions/subdomains. For each subdomain, the unknowns are divided into interior and boundary quantities. The quantities on the boundary are shared between multiple subdomains, and are related via appropriate transmission conditions. Via manipulations, one only solves for unknowns on the boundary of the subdomains which is far fewer that the overall number of unknowns. All interior unknowns can be related to the ones on the boundary. Note, this differs from the domain decomposition approaches currently used for FDTD, which takes a Schwarz approach which uses overlapping subdomains, requiring iteration of each subdomain to get a converged solution\cite{lai2015domain}.

Thus, the main contributions of this paper are as follows: (a) we derive two FETI schemes for time domain Maxwell equation solvers, (b) these methods are then used within an unconditionally stable EM-FEMPIC framework using an exact current mapping scheme, and (c) we present a number of results that demonstrate the efficacy of the proposed technique. 

The rest of this paper is organized as follows. Section \ref{sec:Prelim} will define the problem setup. Section \ref{sec:map} will briefly review particle mapping approach used followed by the field discretization in space and time in Section \ref{sec:fields}. Section \ref{sec:ddm} will define the two domain decomposition formulations. Results using both of these methods will be presented in Section \ref{sec:results} followed by conclusions and future work in Section \ref{sec:conclusions}.

\section{Preliminaries \label{sec:Prelim}}
Let a region $\Omega \in \mathbb{R}^3$ be bounded by surface $\partial\Omega$ and contain at least one charged species.
The charges are represented by a phase space distribution function (PSDF) that is subject to the Maxwell-Vlasov equation
\begin{align} \label{eq:vlasov}
  \partial_t f(t,\vb{r},\vb{v})  + \vb{v} \cdot \nabla f(t,\vb{r},\vb{v}) + \\ \frac{q}{m} [\vb{E}(t,\vb{r}) + \vb{v} \times \vb{B}(t,\vb{r})] \cdot \nabla_v f(t,\vb{r},\vb{v}) = 0. \nonumber
\end{align}
The electric field $\vb{E}(t,\vb{r})$ and magnetic flux density $\vb{B}(t,\vb{r})$ are obtained through the mixed finite element method, which discretizes Faraday's law and Ampere's law.
The PSDF is evolved through the PIC cycle, which consists of:
\begin{enumerate}
    \item Gathering the particles onto the mesh to be mapped into the field solve,
    \item Solving for the fields due to the current from the particle motion,
    \item Interpolating the fields at the particle positions,
    \item Pushing the particles to new positions.
\end{enumerate}
 The following subsections will describe how Maxwell's equations are discretized in space and time, and how the particles are mapped into the field solver.

\subsection{Current Mapping \label{sec:map}}
The charge and current density is defined using the PSDF conventional definition, $\rho(t,\vb{r}) = q \int_\Omega f(t,\vb{r},\vb{v}) d\vb{v}$ and $\vb{J}_\rho(t,\vb{r}) = q\int_\Omega \vb{v}f(t,\vb{r},\vb{v})d\vb{v}$.
The PSDF is approximated by using $N_p$ samples, or particles, with shape $S(\vb{r})$ such that the charge and current density are defined as $\rho(t,\vb{r}) = q\sum_{p=1}^{N_p} S(\vb{r}-\vb{r}_p(t))$ and $\vb{J}_\rho(t,\vb{r}) = q\sum_{p=1}^{N_p} \vb{v}(t)S(\vb{r}-\vb{r}_p(t))$
where $\vb{r}_p(t)$ and $\vb{v}_p(t)$ are the position and velocity of particle $p$. In this work, the shape functions are delta functions, though other shape functions are also valid and can be used to improve noise quality. 
The proper function to measure the charge density is the 0-form and the 1-form for the current density.
Different schemes to evolve Newton's equations can be used to preserve certain quantities like phase.
In this work, to maintain accuracy of the time evolution at larger time step sizes, the Adams-Bashforth scheme is used.
Lastly, the time integral of the current density
\begin{equation}
    \vb{G}(t,\vb{r}) = \int_0^td\tau\vb{J}_\rho (\tau, \vb{r})
\end{equation}
is used to allow the particle current to evolve consistently with the fields.
Otherwise, there is a disconnect between the continuity equation and Gauss' law.

\subsection{Field discretization\label{sec:fields}}

\subsubsection{Spatial Discretization}
The FEM formulation used in this work is the mixed finite element method.
The basis functions used are the 1-form $\vb{W}^{(1)}(\vb{r})\in H(curl;\Omega)$ which allows for tangential continuity of fields, and the 2-form $\vb{W}^{(2)}(\vb{r})\in H(div;\Omega)$ which allows for normal continuity of fluxes. 
The reconstructed fields used to accelerate the particles are defined as $\vb{E}(t,\vb{r}) = \sum_{i=1}^{N_e}e_i(t)\vb{W}^{(1)}(\vb{r})$ and $\vb{B}(t,\vb{r}) = \sum_{i=1}^{N_f}b_i(t)\vb{W}^{(2)}(\vb{r})$ where $N_e$ are the number of edges and $N_f$ are the number of faces.
Testing Faraday's law with the 2-form and Ampere's law with the 1-form leads to the
semidiscrete Maxwell system
\begin{align}\label{eq:maxwell_semi}
\begin{split}
\underbrace{\mqty[ [\star_{\mu^{-1}}]& 0 \\ 0 &[\star_{\epsilon_0}]]}_{\Bar{\Bar{A}}_1}&\mqty[\partial_t \bar{B}\\ \partial_t \bar{E} ]\\
&+ \underbrace{\mqty[0&  [M_c] \\ -c^2[M_c]^T & c[\star_I] ]}_{\Bar{\Bar{A}}_0} \mqty[\bar{B} \\\bar{E} ] = \underbrace{\mqty[0\\ \frac{\bar{L}}{\varepsilon} ]}_{\bar{\bar{F}}}
\end{split}
\end{align}
where the degree of freedom vectors $\bar{E} = [e_1(t),e_2(t),\dots,e_{N_e}(t)]$, $\bar{B} = [b_1(t), b_2(t),\dots,b_{N_f}(t)]$, and $\bar{L}=[l_1(t),l_2(t),...l_{N_e}(t)]$ with $l_i (t) = \langle \vb{W}_i^{(1)} (\vb{r}),\vb{J}_s(t,\vb{r})- \partial_t\vb{G}(t,\vb{r}) \rangle $.
The surface currents $\vb{J}_s(t,\vb{r})$ exist on $\partial\Omega$ due to a Neumann or Robin boundary condition. 
The integrated particle current  is chosen to satisfy charge conservation for arbitrary time marching schemes.

\subsubsection{Temporal Discretization}
This work utilizes the Newmark-$\beta$ time marching scheme to evolve the fields in time with parameters $\gamma$ and $\beta$, which creates an unconditionally stable time marching scheme when $\gamma=.5$ and $\beta=.25$.
Other choices can lead to different stability conditions. 
The parameter choices effect the definition of a temporal basis function $N(t)$ that is tested by a function $W(t)$ \cite{zienkiewicz1977new}.
When \eqref{eq:maxwell_semi} is discretized with Newmark-$\beta$ with $\gamma=0.5$ and $\beta=0.25$, the fully discretized system becomes
 \begin{equation}\label{eq:maxnmb}
 \begin{split}
     &(0.5\bar{\bar{A}}_1 +0.25\Delta_t\bar{\bar{A}}_0)\bar{X}^{n+1} -.5\Delta_t\bar{\bar{A}}_0\bar{X}^n\\
     &+(0.5\bar{\bar{A}}_1 +.25\Delta_t\bar{\bar{A}}_0)\bar{X}^{n-1} +0.5\Delta_t\tilde{G}^{n + 1} + 0.5\Delta_t\tilde{G}^{n-1}\\
     &+0.25\Delta_t\tilde{F}^{n+1}-0.5\Delta_t\tilde{F}^n+0.25\tilde{F}^{n-1}=0.
 \end{split}
 \end{equation}
Here, $\bar{X}^m=[\bar{B}^m\; \bar{E}^m]^T$, $\tilde{G}^m = [0\; \varepsilon^{-1}\bar{G}]^{m,T}$, and $\tilde{F}^m = [0\; \varepsilon^{-1}\bar{J}^m_s]$ at time step $m$.
The boundary current $\bar{J}^m_s$ is tangential to a surface. As it has no normal component, it does not contribute to Gauss' law and therefore does violate charge conservation.

\section{Domain Decomposition \label{sec:ddm}}

\begin{figure}
\centering
\includegraphics[scale=.3]{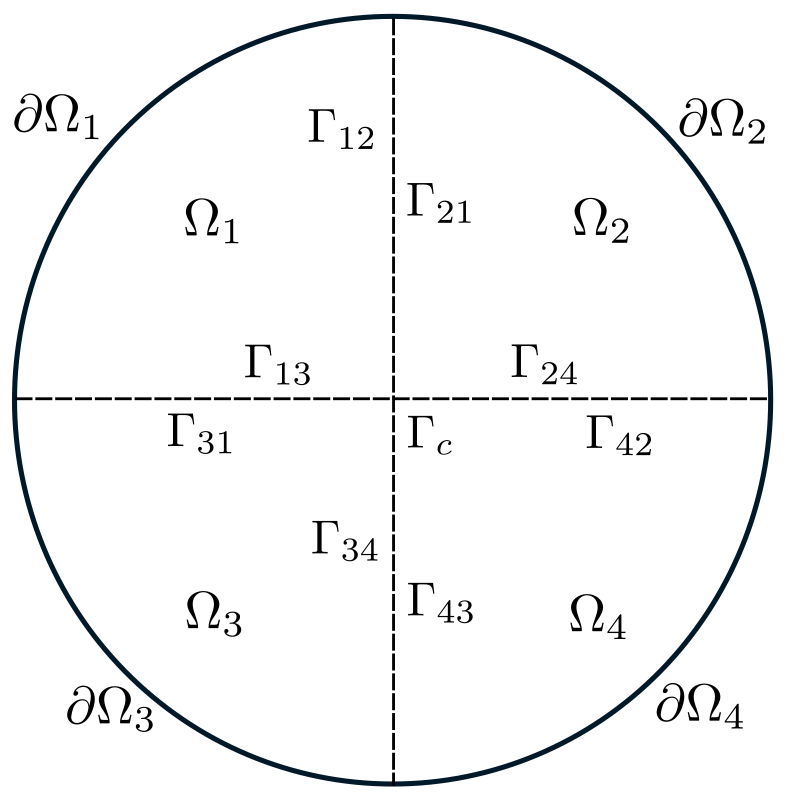}
\caption{Region $\Omega$ partitioned into $N_v$ subdomains. }
\label{fig:ddmfig}
\end{figure}

Consider the region $\Omega$ depicted in Fig. \ref{fig:ddmfig} that is divided into $N_v$ non-overlapping subdomains where subdomain $\Omega_i$ and $\Omega_j$ are separated by boundary $\Gamma_{ij}$. The junction of more than two volumes is a "corner" denoted by $\Gamma_c$.
In each subdomain $\Omega_i$, \eqref{eq:maxnmb} is defined and is assumed to be a self contained ``primal" problem that has fictitious excitations from "dual" unknowns.
The dual unknowns, the Lagrange multipliers denoted by $\Lambda$, effect either Neumann or Robin boundary conditions such that the equivalence theorem can be applied for each $\Omega_i$ with respect to $\Omega$ using the external currents on $\partial\Omega_i$, the fictitious boundary current from $\Lambda$, and the particle current in $\Omega_i$. 
By solving for the dual unknowns and a small subset of the primal unknowns, the overall computational cost of solving the original monolithic system in \eqref{eq:maxnmb} is reduced.
\subsection{MFEM-DP 1}
The first formulation presented uses a Neumann-like boundary condition to enforce the continuity conditions of the fields at the interior boundaries,
\begin{subequations}
\begin{equation}\label{eq:bce}
    \hat{n}\cross\vb{E}_i(t,\vb{r}) = \hat{n}\cross\vb{E}_j(t,\vb{r}) \;\; \vb{r}\in \Gamma_{ij}
\end{equation}
\begin{equation}\label{eq:bcb}
    \hat{n}\cdot\vb{B}_i(t,\vb{r}) = \hat{n}\cdot\vb{B}_j(t,\vb{r})\;\; \vb{r}\in \Gamma_{ij}.
\end{equation}
\end{subequations}
This is accomplished by ensuring $\vb{x}^n_i = \vb{x}^n_j |_{\Gamma_{ij}}$ and defining 
\begin{equation}\label{eq:F1cur}
    \hat{n}\cross\vb{H}_i(t,\vb{r}) = \Lambda(t,\vb{r})\;\; \vb{r}\in \Gamma_{ij}.
\end{equation}
The system of equations to be solved in each sub-region
\begin{equation}\label{eq:simpleF1}
    \left(0.5\dbar{A}^{(i)}_1+0.25\Delta_t\dbar{A}^{(i)}_0\right)\bar{X}^{(i),n+1} = \mathcal{L}^{(i)}-\Lambda^{(i)} 
\end{equation}
where
\begin{equation}
\begin{split}
    &\mathcal{L}^{(i)} = 0.5\Delta_t\bar{\bar{A}}_0\bar{X}^{(i),n}\\
     &-\left(0.5\bar{\bar{A}}^{(i)}_1 +.25\Delta_t\bar{\bar{A}}^{(i)}_0\right)\bar{X}^{(i),n-1} -0.5\Delta_t\tilde{G}^{(i),n + 1} \\
     &- 0.5\Delta_t\tilde{G}^{(i),n-1}-0.25\Delta_t\tilde{F}^{(i),n+1}+0.5\Delta_t\tilde{F}^{(i),n}\\
     & -0.25\tilde{F}^{(i),n-1}.
\end{split}
\end{equation}
It is noted that the time history of the Lagrange multiplier is not explicitly stored, unlike the other currents in the system. Let the unknowns defined on $\Gamma_c$ be denoted by the subscript $c$ and all other unknowns in $\Omega_i$ denoted by subscript $r$, such that \eqref{eq:simpleF1} can be written as
\begin{equation}\label{eq:blockF1}
    \begin{bmatrix}
    \dbar{A}^{(i)}_{rr} & \dbar{A}^{(i)}_{rc}\\
    \dbar{A}^{(i)}_{cr} & \dbar{A}^{(i)}_{cc}
    \end{bmatrix}
    \begin{bmatrix}
    \bar{X}^{(i),n+1}_r\\
    \bar{X}^{(i),n+1}_c
    \end{bmatrix}=
    \begin{bmatrix}
    \mathcal{L}^{(i)}_r\\
    \mathcal{L}^{(i)}_c
    \end{bmatrix}-
    \begin{bmatrix}
    \bar{\Lambda}^{(i)}_r\\
    \bar{\Lambda}^{(i)}_c
    \end{bmatrix}
\end{equation}
where $\dbar{A}=.5\dbar{A}_1+.25\Delta_t\dbar{A}_0$.
The first set of equations can be rewritten as
\begin{equation}\label{eq:rF1}
    \bar{X}^{(i),n+1}_r = \dbar{A}^{(i),-1}_{rr}(\mathcal{L}^{(i)}_r - \dbar{A}^{(i),-1}_{rc}\bar{X}^{(i),n+1}_c - \bar{\Lambda}^{(i)}_r)
\end{equation}
which can be substituted into the second set of equations
\begin{equation}\label{eq:cF1}
\begin{split}
\Big( \dbar{A}^{(i)}_{cc}- \dbar{A}^{(i),-1}_{cr} &\dbar{A}^{(i),-1}_{rr}\dbar{A}^{(i)}_{rc}\Big) \bar{X}^{(i),n+1}_c =\mathcal{L}^{(i)}_c\\
    &- \dbar{A}^{(i)}_{cr}\dbar{A}^{(i),-1}_{rr}\left(\mathcal{L}^{(i)}_r - \bar{\Lambda}^{(i)}_r\right) -\bar{\Lambda}^{(i)}_c.
\end{split}
\end{equation}
    
Define a signed Boolean matrix $\dbar{B}_r^{(i)}$ that maps unknowns in $\Omega_i$, excluding those on $\Gamma_c$, to a set of unique global unknowns defined on $\cup\Gamma_{ij}$.
The definition of $\dbar{B}_r^{(i)}$ is such that $\sum_{i=1}^{N_v}\dbar{B}_r^{(i)}\bar{x}^{(i)}=0$, which enforces \eqref{eq:bce} and \eqref{eq:bcb}.
Furthermore, the transpose $\dbar{B}_r^{(i),T}$ forms a map from the global set of boundary unknowns to those of a particular volume.
A similar unsigned Boolean matrix can be defined for unknowns that lie on $\Gamma_c$ within $\Omega_i$, $\dbar{B}_c$, which maps to a set of unique global unknowns.
Using $\dbar{B}_r$ on \eqref{eq:rF1} and summing over all volumes results in
\begin{equation}\label{eq:F1lr}
    \dbar{K}_{rr}\Lambda_r = \bar{\mathcal{L}}_r - \dbar{K}_{rc}\bar{X}_c
\end{equation}
where 
\begin{subequations}
\begin{equation}
    \dbar{K}_{rr} = \sum_{i=1}^{N_v}\dbar{B}_r\dbar{A}^{(i)}_{rr}\dbar{B}_r^{(i),T}
\end{equation}
\begin{equation}
    \dbar{K}_{rr}= \sum_{i=1}^{N_v}\dbar{B}_r\dbar{A}^{(i)}_{rr}\dbar{A}^{(i)}_{rc}\dbar{B}_c^{(i),T}
\end{equation}
\begin{equation}
    \bar{\mathcal{L}}_r = \sum_{i=1}^{N_v} \dbar{B}_r^{(i)}\dbar{A}^{(i),-1}_{rr}\mathcal{L}^{(i)}_r
\end{equation}
\end{subequations}
and $\bar{X}_c$ the degree of freedom vector for corner unknowns.

The boundary condition \eqref{eq:bce} is enforced at the corners of $\Omega_i$ as $\dbar{B}_c^{(i)}$ is defined so that $\sum_{i=1}^{N_v}\dbar{B}_c^{(i),T}\lambda^{(i)}_{c} = 0$.
Using that definition, \eqref{eq:cF1} becomes
\begin{equation}\label{eq:F1uc}
    \dbar{K}_{cc}\bar{X}^{n+1}_c = \bar{\mathcal{L}}_c + \dbar{K}_{cr}\Lambda_r.
\end{equation}
where
\begin{subequations}
\begin{equation}
  \dbar{K}_{cc} = \sum_{i=1}^{N_v}\dbar{B}_c^{(i)} (\dbar{A}^{(i)}_{cc}- \dbar{A}^{(i),-1}_{cr}\dbar{A}^{(i),-1}_{rr}\dbar{A}^{(i),-1}_{rc})\dbar{B}_c^{(i),T}  
\end{equation}
\begin{equation}
  \dbar{K}_{cr} = \sum_{i=1}^{N_v}\dbar{B}_c^{(i)}\dbar{A}^{(i)}_{cr}\dbar{A}^{(i),-1}_{rr}\dbar{B}_r^{(i),T}  
\end{equation}
\begin{equation}
    \bar{\mathcal{L}}_c = \sum_{i=1}^{N_v} \dbar{B}_c^{(i)}(\mathcal{L}^{(i)}_c - \dbar{A}^{(i)}_{cr}\dbar{A}^{(i),-1}_{rr}\mathcal{L}^{(i)}_r).
\end{equation}
\end{subequations}
As in the vector wave equation domain decomposition formulations, $\dbar{K}_{cc}$ is sparse and generally much smaller than $\dbar{K}_{rr}$.
Therefore, if solving the boundary problem directly, it is possible to define
\begin{equation}
    \bar{X}^{n+1}_c = \dbar{K}^{-1}_{cc}\left(\bar{\mathcal{L}}_c + \dbar{K}_{cr}\Lambda_r\right)
\end{equation}
and then use substitution into \eqref{eq:F1lr} to yield
\begin{equation}\label{eq:f1schur}
  \dbar{\mathcal{K}}_{rr}\Lambda_r = \bar{\mathcal{L}}_r - \dbar{K}_{rc}\dbar{K}_{cc}^{-1}\bar{\mathcal{L}}_c
\end{equation}
where $\dbar{\mathcal{K}}_{rr}= \left(\dbar{K}_{rr}+\dbar{K}_{rc}\dbar{K}_{cc}^{-1}\dbar{K}_{cr}\right)$.
If using an iterative solver, \eqref{eq:F1lr} and \eqref{eq:F1uc} can be solved as a coupled system instead.

\subsection{MFETI-DP2}

In the previous formulation, a unique Lagrange multiplier was defined for each edge and face on $\Gamma$. 
This formulation enforces field continuity though a Neumann-like boundary condition.
A more robust enforcement of field continuity is to use a Robin boundary condition \cite{jin2015finite}. The boundaries between $\Omega_i$ and $\Omega_j$, $\Gamma_{ij}$ and $\Gamma_{ji}$, are distinct and have a unique Lagrange multiplier defined for each subdomain. Instead of \eqref{eq:F1cur}, the boundary condition on the Lagrange multipliers becomes
\begin{equation}\label{eq:F2cur}
    \Lambda^{(i)}(t,\vb{r}) = \hat{n}_i \cross \vb{H}(t,\vb{r}) - \eta^{-1}\hat{n}_i\cross\hat{n}_i\cross\vb{E}(t,\vb{r}) \in \Gamma_{ij}.
\end{equation}
The edges on $\Gamma_{ij}$, as well as the interior unknowns and unknowns on $\partial\Omega_i$ are included in a vector $\bar{X}^{(i)}_I$.
The Lagrange multipliers for the corner edges remain the same, with a unique Lagrange multiplier for each feature.
The faces on an $\Gamma_{ij}$ are grouped together with the corner edges, which we will denote as $\bar{X}_N$. 
The system of equations solved in each subdomain is
\begin{equation}\label{eq:blockF2}
    \begin{bmatrix}
    \dbar{A}^{(i)}_{II} & \dbar{A}^{(i)}_{IN}\\
    \dbar{A}^{(i)}_{NI} & \dbar{A}^{(i)}_{NN}
    \end{bmatrix}
    \begin{bmatrix}
    \bar{X}^{(i),n+1}_I\\
    \bar{X}^{(i),n+1}_N
    \end{bmatrix}=
    \begin{bmatrix}
    \mathcal{R}^{(i)}_I\\
    \mathcal{R}^{(i)}_N
    \end{bmatrix}-
    \begin{bmatrix}
    .5\Delta_t\bar{\Lambda}^{(i),n+1}_I\\
    \bar{\Lambda}^{(i)}_N
    \end{bmatrix}\in \Omega_i
\end{equation}
where given
\begin{equation}
\begin{split}
    &\mathcal{R} = .5\Delta_t\bar{\bar{A}}_0\bar{X}^n\\
     &-(.5\bar{\bar{A}}_1 +.25\Delta_t\bar{\bar{A}}_0)\bar{X}^{n-1} -.5\Delta_t\tilde{G}^{n + 1} - .5\Delta_t\tilde{G}^{n-1}\\
     &-.25\Delta_t\tilde{F}^{n+1}+.5\Delta_t\tilde{F}^n-.25\tilde{F}^{n-1}-.5\Delta_t\Lambda^{n}+.25\Lambda^{n-1},
\end{split}
\end{equation}
$\mathcal{R}^{(i)}_I$ contains $\mathcal{R}$ associated with interior unknowns, external unknowns, and unknowns of edges on $\Gamma_{ij}$ and $\mathcal{R}^{(i)}_N$ contain the remaining unknowns in subdomain $\Omega_i$.
To derive the set of equations to solve for the dual unknowns, first consider the summation of \eqref{eq:F2cur} for two volumes
\begin{equation}\label{eq:sumF2cur}
    \Lambda^{(i)}(t,\vb{r}) + \Lambda^{(j)}(t,\vb{r}) +(Y_i+Y_j)\hat{n}_i\cross\hat{n}_i\cross\vb{E}(t,\vb{r}).
\end{equation}
Define an unsigned Boolean matrix such that $\dbar{B}_I^{(i)}\bar{X}_I = \bar{e}_I^{(i)}$, the unknowns for the electric field on $\Gamma_{ij}$.
Also, define the map from the boundary edges on $\Gamma_{ij}$ to its counterpart on $\Gamma_{ji}$, $\bar{T}_{ij}$.
Testing \eqref{eq:sumF2cur} with curl-conforming basis functions $\vb{W}^{e}$, generalizing the case to any number of subdomains, and using $\dbar{B}_I^{(i)}$ and $\dbar{T}_{ji}$ restrict and map the unknowns of $\Omega_j$ to $\Omega_i$ yields
\begin{equation}\label{eq:semiF2cur}
    \Lambda^{(i)} + \sum_{j \in \sigma(i)}\dbar{T}_{ij}(\Lambda^{(j)} - (Y_i+Y_j)\dbar{M}_{T}^{(j)}\dbar{B}^{(j)}_I\bar{X}^{(j)}_I) = 0
\end{equation}
where $\sigma(i)$ are all subdomains $\Omega_j$ which share a boundary with $\Omega_i$ and the matrix   $\dbar{M}_T^{(i)} \langle \hat{n}\times\vb{W}^{e},\hat{n}\times\vb{W}^{e}\rangle |_{\Gamma_{ij}}$.

Rewrite the first set of equations in \eqref{eq:blockF2} as
\begin{equation}\label{eq:IF2}
    \bar{X}^{(i),n+1}_I = \dbar{A}^{(i),-1}_{II}(\mathcal{R}^{(i)}_I -  .5\Delta_t\bar{\Lambda}^{(i),n+1}_I - \dbar{A}^{(i)}_{IN}\bar{X}^{(i),n+1}_N)
\end{equation}
Now, \eqref{eq:IF2} can be plugged into \eqref{eq:semiF2cur} to yield
\begin{equation}\label{eq:f2imp1}
\begin{split}
    \Lambda^{(i)}\!\! &+ \!\!\!\!\!\sum_{j\in \sigma(i)}\!\!T_{ij}(\dbar{I}^{(j)} - (Y_i+Y_j)\dbar{M}_T^{(j)}\dbar{B}_I^{(j)}\dbar{A}_{II}^{(j),-1}\dbar{B}_I^{(j),T})\Lambda^{(j)}\\
    &= \sum_{j\in \sigma(i)}T_{ij}\dbar{B}_I^{(j)}\dbar{A}_{II}^{(j),-1}(\dbar{A}_{IN}^{(j)}\dbar{B}_N^{(j)}\bar{X}_N - \mathcal{R}^{(j)}_I).
\end{split}
\end{equation}
In order to complete the definition of the equation to solve for $\Lambda_I$ over $\cup\Gamma_{ij}$, we first define the equation for $\bar{X}_N$, the unknowns associated with faces on $\Gamma_{ij}$ and edges on $\Gamma_c$.

The derivation of the equation to solve for $\bar{X}_N$ follows a similar path as $\bar{X}_c$ from MFET-DP1. First, substitute \eqref{eq:IF2} into the second set of equations in \eqref{eq:semiF2cur} to yield
\begin{equation}
\begin{split}
\Big( \dbar{A}^{(i)}_{NN}- \dbar{A}^{(i),-1}_{NI} &\dbar{A}^{(i),-1}_{II}\dbar{A}^{(i)}_{IN}\Big) \bar{X}^{(i),n+1}_N =\mathcal{R}^{(i)}_N\\
    &- \dbar{A}^{(i)}_{NI}\dbar{A}^{(i),-1}_{II}\left(\mathcal{R}^{(i)}_I - \bar{\Lambda}^{(i)}_I\right) -\bar{\Lambda}^{(i)}_N.
\end{split}
\end{equation}
The Boolean matrix $\dbar{B}_N^{(i)}$ is defined as
\begin{equation}
    \dbar{B}_N^{(i)} = \begin{bmatrix}
    \dbar{B}_{r,f}^{(i)} & 0\\
    0 & \dbar{B}_{c}^{(i)}
    \end{bmatrix}
\end{equation}
to enforce \eqref{eq:bcb} for the faces on $\Omega_i$ and \eqref{eq:bce} edges on corners where $\dbar{B}_{r,f}^{(i)}$ is the portion of $\dbar{B}_{r}^{(i)}$ that acts on the face degrees of freedom.
Using a similar process as \eqref{eq:F1uc}, the equation for 
\begin{equation}\label{eq:F2un}
    \dbar{K}_{NN}\bar{X}^{n+1}_N = \bar{\mathcal{R}}_N + \dbar{K}_{NI}\Lambda_I.
\end{equation}
where
\begin{subequations}
\begin{equation}
  \dbar{K}_{NN} = \sum_{i=1}^{N_v}\dbar{B}_N^{(i)} (\dbar{A}^{(i)}_{NN}- \dbar{A}^{(i),}_{NI}\dbar{A}^{(i),-1}_{II}\dbar{A}^{(i),-1}_{IN})\dbar{B}_N^{(i),T}  
\end{equation}
\begin{equation}
  \dbar{K}_{NI} = \sum_{i=1}^{N_v}\dbar{B}_N^{(i)}\dbar{A}^{(i)}_{NI}\dbar{A}^{(i),-1}_{II}\dbar{B}_I^{(i),T}\dbar{B}_L^{(i),T}
\end{equation}
\begin{equation}
    \bar{\mathcal{R}}_N = \sum_{i=1}^{N_v} \dbar{B}_N^{(i)}(\mathcal{R}^{(i)}_N - \dbar{A}^{(i)}_{NI}\dbar{A}^{(i),-1}_{II}\mathcal{R}^{(i)}_I).
\end{equation}
\end{subequations}

The global system of equations for the edge Lagrange multipliers
Define a Boolean matrix $\dbar{B}_L^{(i)}$ that maps the local Lagrange multipliers to a global vector.
The Boolean matrix is applied to \eqref{eq:f2imp1} and discretized in time to define the equation
\begin{equation}\label{eq:f2lam}
    \dbar{K}_{II}\Lambda_I = \bar{\mathcal{R}}_I -\dbar{K}_{IN}\bar{N}^{n+1}_N
\end{equation}
where
\begin{subequations}
\begin{equation}
\begin{split}
    \dbar{K}_{II} &= \underline{I} +\sum_{i=1}^{N_v}  \underline{B}^{(i)}_L\\
    &\sum_{j \in \sigma(i) } \dbar{T}_{ij}(\dbar{I}^{(j)} - \underline{M}_T^{(j)}\dbar{B}_I^{(j)}\dbar{A}_{II}^{(j),-1})\dbar{B}_I^{(j),T}\underline{B}_L^{(j),T}
\end{split}
\end{equation}
\begin{equation}
    \dbar{K}_{IN} = -\sum_{i=1}^{N_v}\underline{B}_L^{(i)}\sum_{j\in\sigma(i)} \underline{T}_{ji}\underline{M}_T^{(j)}\underline{B}_I^{(j)}\dbar{A}_{II}^{(j),-1}\dbar{A}^{(j)}_{IN}\underline{B}_N^{(j),T} 
\end{equation}
\begin{equation}
    \bar{\mathcal{R}}_{I} = -\sum_{i=1}^{N_v} \underline{B}_L^{(j)}\sum_{j\in\sigma(i)}\underline{T}_{ji}\dbar{M}^{(j)}_T\underline{B}^{(j)}_{I}\dbar{A}_{II}^{(j),-1}\mathcal{R}^{(i)}_I)
\end{equation}
\end{subequations}
It can be solved as a coupled system with \eqref{eq:F2un} or in a Schur complement system similar to \eqref{eq:f1schur} using
\begin{equation}\label{eq:f2schur}
    \dbar{\mathcal{K}}_{II}\Lambda_I = \bar{\mathcal{L}}_I - \dbar{K}_{IN}\dbar{K}_{NN}^{-1}\bar{\mathcal{L}}_N.
\end{equation}
where $\dbar{\mathcal{K}}_{II} = \left(\dbar{K}_{II}-\dbar{K}_{IN}\dbar{K}_{NN}^{-1}\dbar{K}_{NI}\right)$.
However, $\dbar{K}_{NN}$ is much larger than its MFETI-DP1 counterpart $\dbar{K}_{cc}$ and will take more time and memory to compute and store its inverse.

\subsection{Discussion}

The proposed domain decomposition formulations reduces the cost of solving for the fields by defining a set of equations on fictitious interior boundaries of the true problem. The definitions of these equations uses small matrix inverses, and therefore are generally not sparse.
Furthermore, those matrix inverses have to be stored in memory.
At the very least, one matrix inverse for each subdomain is needed, making the memory storage more than solving for the original system.
However, it is expected that for direct inverses, inverting either the coupled equations of \eqref{eq:F1lr} and \eqref{eq:F1uc} or \eqref{eq:f2lam} and \eqref{eq:F2un} would be more efficient than solving the original problem.
If storing the inverse of the matrix associated with $\Gamma_c$ is possible, \eqref{eq:f1schur} and \eqref{eq:f2schur} are even smaller in size.
The following discussion uses a sample problem to further probe properties of MFETI-DP1 and MFETI-DP2. In particular, we examine the performance of the method with a sample geometry with several configurations of subdivision.

\begin{figure}
\centering
\includegraphics[width=\linewidth]{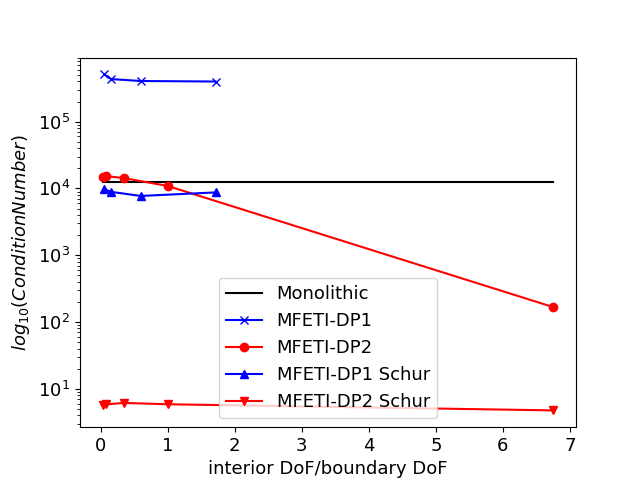}
\caption{ Condition number of system matrix. }
\label{fig:cond}
\end{figure}

\begin{table}[]
    \centering
        \caption{Average number of unknowns per volume and ratio of interior to boundary unknowns for different decompositions of $\Omega_{rect}$.}    \label{tab:rect}
    \begin{tabular}{c|c|c|c|c}
    &\multicolumn{2}{c}{MFETI-DP1} &\multicolumn{2}{c}{MFETI-DP2}\\
    $N_v$ & avg unk. & $\alpha$ & avg unk. &$\alpha$\\
    \hline
         2 & 1815 & N/A & 1755 & 6.75 \\
         5 & 776 & 1.71 & 705 & 1.00\\
         10 & 404 & 0.60 & 351 & 0.34 \\
         25 & 170 & 0.15 & 133 & 0.08 \\
         50 & 89 & 0.05 & 60 & 0.03 \\
    \end{tabular}
\end{table}
Define $\Omega_{rect}$ as a rectangular region in free space with dimension .25 m $\times$ .2 m $\times$ .3 m discretized with average edge length of 5.5 cm. 
The region is the split into subdomains in several configurations listed in Table \ref{tab:rect} so that MFETI-DP1 and MFETI-DP2 can be applied, where $\alpha$ is the ratio of interior unknowns to the number of boundary unknowns.
This table is provided to provide the connection between the number of subdomains and $\alpha$ for the test example, as $\alpha$ will be more extensible to other geometries than a set number of subdomains. 
Figure \ref{fig:cond} shows the condition number of the matrix formed by the coupled equations \eqref{eq:F1lr} and \eqref{eq:F1uc} for MFETI-DP1, \eqref{eq:f2lam} and \eqref{eq:F2un} for MFETI-DP2, as well as the Schur complement forms $\dbar{\mathcal{K}}_{rr}$ and $\dbar{\mathcal{K}}_{II}$ with different numbers of subdivisions compared to the condition number of  of the monolithic system matrix $\dbar{A}$.
Solving the coupled equations results in condition number that are generally at or worse than the monolithic solve.
Using the Schur complement results in condition numbers comparable to the monolithic solve for MFETI-DP1 and several orders of magnitude smaller for MFETI-DP2.

\begin{figure}
\centering
\includegraphics[width=\linewidth]{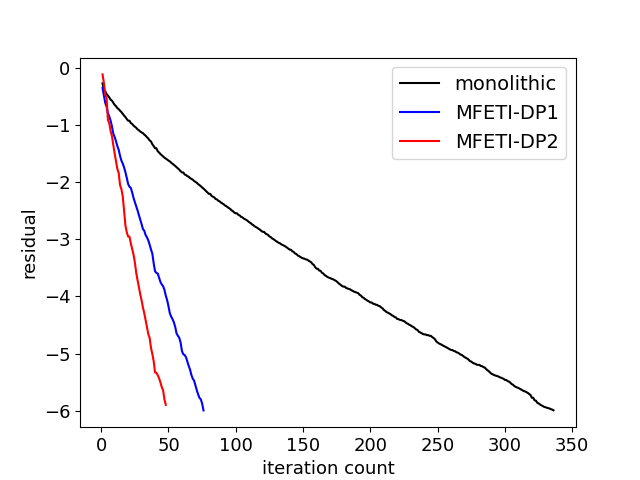}
\caption{ Convergence of history of GMRES for monolithic, MFETI-DP1, and MFETI-DP2. }
\label{fig:fetiIter}
\end{figure}

Typically, MFETI-DP1 and MFETI-DP2 when used with iterative solvers  converge faster than the monolithic solve.
Consider a domain $\Omega_{rect}$ is illuminated by an incident field defined by
\begin{equation}\label{eq:excite}
\bar{E}(\bar{r},t) = \hat{y}\cos(2\pi f_0 \tau)e^{-(\tau  - 8\sigma)^2/2\sigma^2}\, \text{({V/m})}.
\end{equation}
The bandwidth of the modulated Gaussian is $\sigma =3(\pi f_{bw} $, where the bandwidth  $f_{bw} = f_{max}-f_0$ with maximum frequency $f_{max} =500$MHz and center frequency $f_0 =300$MHz. The time step size is $0.333$ ns which is ten times larger than what could be taken by a similar FDTD solver or leapfrog method for MFEM.
Figure \ref{fig:fetiIter} shows the iteration counts for the coupled solves for MFETI-DP1 and MFETI-DP2 compared to the monolithic solve using GMRES with a tolerance of $10^{-6}$ and restarted after 20 iterations when $\Omega_{rect}$ is illuminated by an incident field.
With a diagonal preconditioner, both domain decomposition approaches perform better than the monolithic solve, with MFETI-DP2 performing best overall.

\begin{figure}
\begin{subfigure}{.85\linewidth}
\centering
\includegraphics[width=\linewidth]{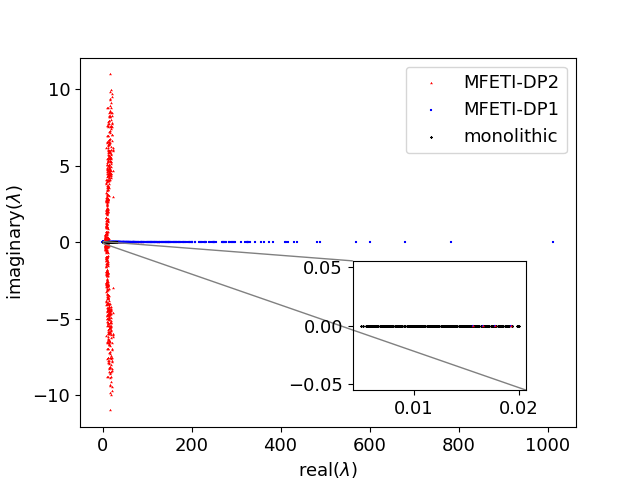}
\caption{MFETI-DP matrix eigenvalues.}\label{fig:beigs}
\end{subfigure}
\begin{subfigure}{.85\linewidth}
\centering
\includegraphics[width=\linewidth]{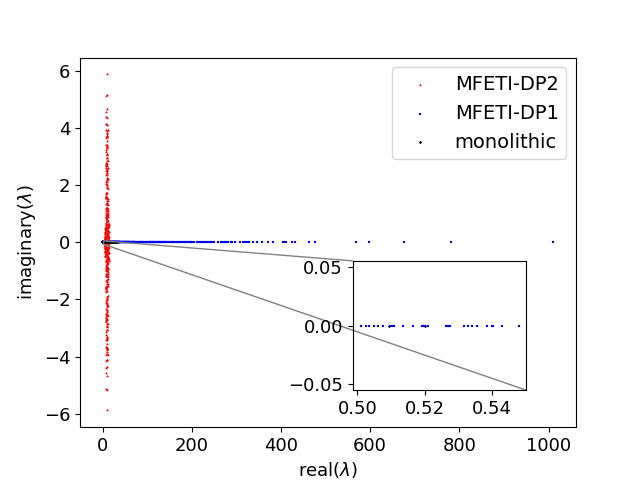}
\caption{MFETI-DP matrix eigenvalues using Schur complement.}\label{fig:seigs}
\end{subfigure}
\caption{ System matrix eigenvalues. }
\label{fig:eigs}
\end{figure}

Next, as alluded to earlier, the principal advantage of Newmark-$\beta$ schemes is the unconditional stability. As a result, it is important to verify that the both MFETI-DP1 and MFETI-DP2 retain this feature. To ascertain this, we examine the eigenspectra of these formulations to understand the stability of the time marching scheme.
For the time marching scheme used in this work to be stable, the matrix used to evolve the system in time must have eigenvalues $\Re(\lambda) >=0$ for the system to be unconditionally stable.
Figure \ref{fig:eigs} shows that for solving both the coupled system and using the Schur complement approach, the domain decomposition methods will be stable.
From the insert in Figure \ref{fig:eigs}, the effect of using the schur complement shifts the eigenvalues of $\dbar{\mathcal{K}}_{rr}$ and $\dbar{\mathcal{K}}_{II}$ away from the imaginary axis, though MFETI-DP2 is shifted further away than MFETI-DP1.


\section{Results\label{sec:results}}

In this section, we present numerical results that show how the proposed algorithms perform compared to a non-DDM approach. Additionally, we will demonstrate that as the fields are almost identical to machine precision, the overall trajectory of the particles are the identical to machine precision as well, i.e., the approach does not induce additional error.

\subsection{Field Comparison}

\begin{figure}
\begin{subfigure}{.85\linewidth}
\includegraphics[width=\linewidth]{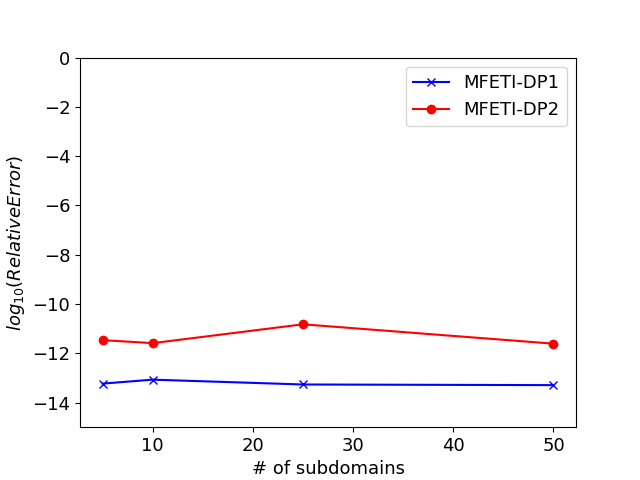}
\caption{ Geometry with average edge length $h_1$. }\label{fig:errH1}
\end{subfigure}
\begin{subfigure}{.85\linewidth}
\includegraphics[width=\linewidth]{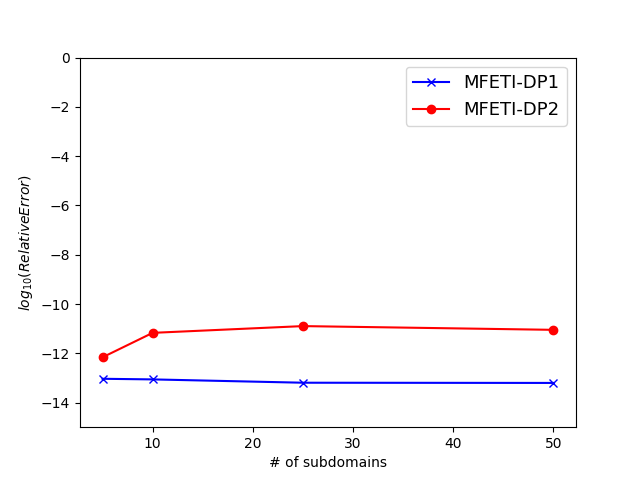}
\caption{ Geometry with average edge length $h_2$. }
\label{fig:errH2}
\end{subfigure}
\begin{subfigure}{.85\linewidth}
\includegraphics[width=\linewidth]{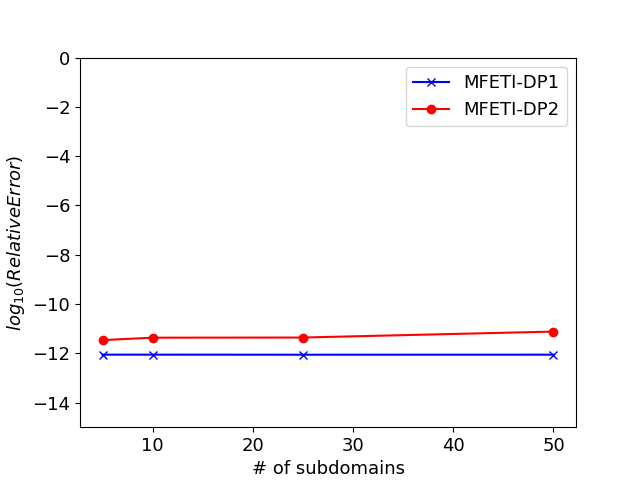}
\caption{ Geometry with average edge length $h_3$. }
\label{fig:errH3}
\end{subfigure}
\caption{ Relative error in electric field for MFETI-DP. }\label{fig:ddmErr}
\end{figure}

The first example shows performance of the methods and difference in the fields.
In this example, a $0.25 m \times 0.2 m \times 1 m$ box is illuminated an incident field defined by
\begin{equation}\label{eq:excite}
\bar{E}(\bar{r},t) = \hat{y}\cos(2\pi f_0 \tau)e^{-(\tau  - 8\sigma)^2/2\sigma^2}\, \text{({V/m})}.
\end{equation}
The bandwidth of the modulated Gaussian is $\sigma =3(\pi f_{bw} $, where the bandwidth  $f_{bw} = f_{max}-f_0$ with maximum frequency $f_{max} =595$MHz and center frequency $f_0 =295$MHz.
The time step size $\delta_t = (150f_0)^{-1}$ and the experiment is run for 1500 time steps.
The boundary conditions on the external boundaries are robin boundary conditions.
Three approaches are compared: monolithic (no DDM), MFETI-DP1, and MFETI-DP2 using three levels of discretization with average edge lengths of $h_1=5.20$cm, $h_2=3.63$cm, and $h_3=2.59$cm.
The DDM meshes were split using METIS, such that no volumes have a regular shape.
Figure \ref{fig:ddmErr}, shows that the fields obtained through domain decomposition for edge lengths $h_1$, $h_2$, and $h_3$ respectively. 
Regardless of the number of subdivisions, the methods remain accurate to near machine precision with respect to the monolithic solve.
This implies that using either domain decomposition formulation would cause the same acceleration on particles as the fields that are similar to each other within machine precision.

\begin{figure}
\centering
\begin{subfigure}{.85\linewidth}
\includegraphics[width=\linewidth]{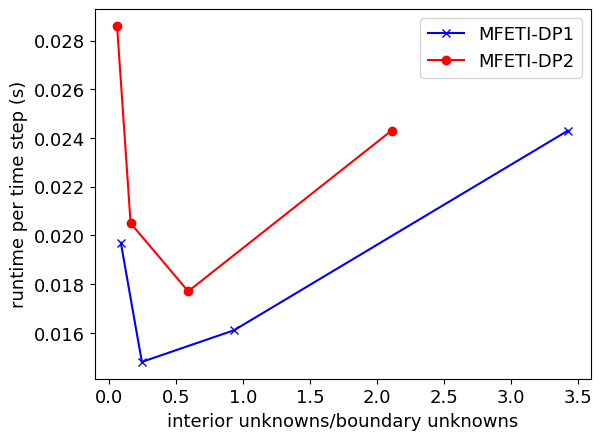}
\caption{ Geometry with average edge length $h_1$. }
\label{fig:timH1}
\end{subfigure}
\begin{subfigure}{.85\linewidth}
\centering
\includegraphics[width=\linewidth]{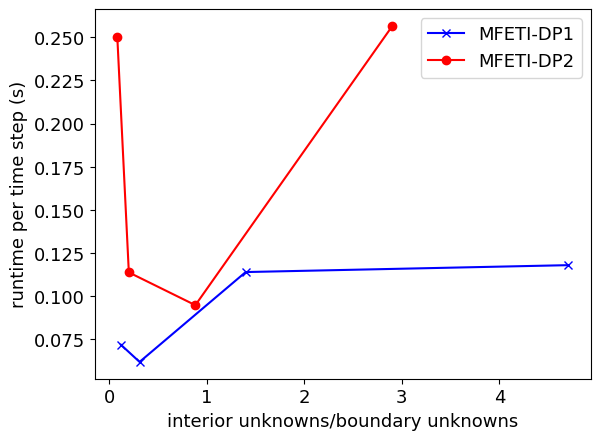}
\caption{ Geometry with average edge length $h_2$. }
\label{fig:timH2}
\end{subfigure}
\begin{subfigure}{.85\linewidth}
\includegraphics[width=\linewidth]{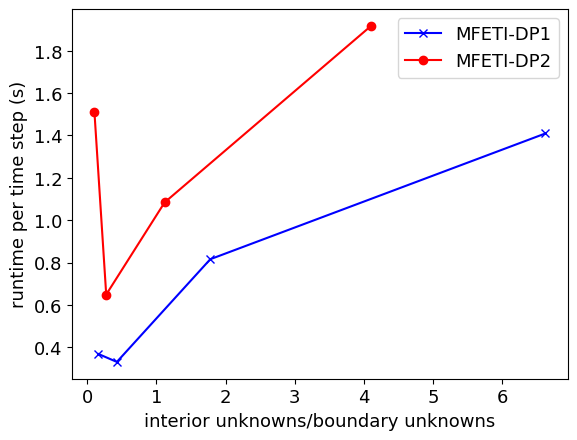}
\caption{ Geometry with average edge length $h_3$. }
\label{fig:timH3}
\end{subfigure}
\caption{ Timing data for MFETI-DP. }\label{fig:ddmTim}
\end{figure}

The next result demonstrates the effect of MFETI-DP on simulation time.
As seen in Figure \ref{fig:ddmTim}, both formulations are faster than the original monolithic solve for any number of subdivisions.
However, there is an optimal ratio of interior volume unknowns to boundary unknowns for the fastest runtime.
For each case, there is a crossover point at which further subdividing $\Omega$ incurs a cost in solving the boundary problem that is greater than the savings by reducing the unknowns in each volume.

\begin{table}[]
    \centering
        \caption{Ratio of interior to boundary unknowns for different decompositions for particle beam test.}    \label{tab:beam}
    \begin{tabular}{c|c|c|c|c}
    &\multicolumn{2}{c}{MFETI-DP1} &\multicolumn{2}{c}{MFETI-DP2}\\
    $N_v$  & $\alpha$& runtime (s) & $\alpha$& runtime (s)\\
    \hline
         5  & 3.000 & 571.4 & 1.893 & 694.1 \\
         10 & 0.955 & 614.8 & 0.652 & 711.9 \\
         25 & 0.257 & 466.7 & 0.181 & 488.6 \\
         50 & 0.094 & 562.5 & 0.077 & 507.6 \\
    \end{tabular}
\end{table}

\subsection{Particle Beam}

\begin{figure}
\centering
\begin{subfigure}{.85\linewidth}
\centering
\includegraphics[width=\linewidth]{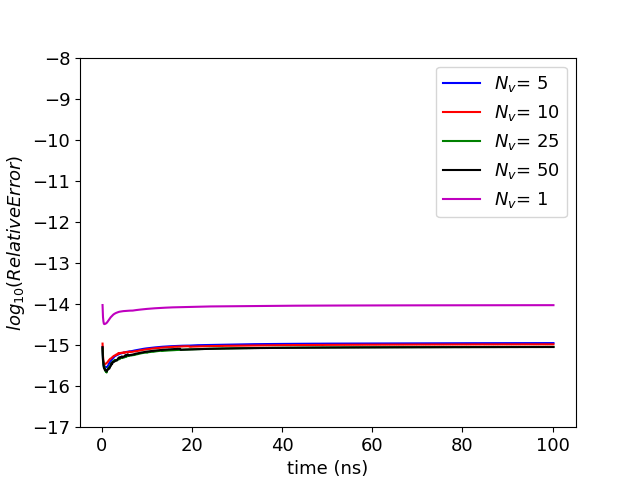}
\caption{ MFETI-DP1 }
\label{fig:dglF1}
\end{subfigure}
\begin{subfigure}{.85\linewidth}
\centering
\includegraphics[width=\linewidth]{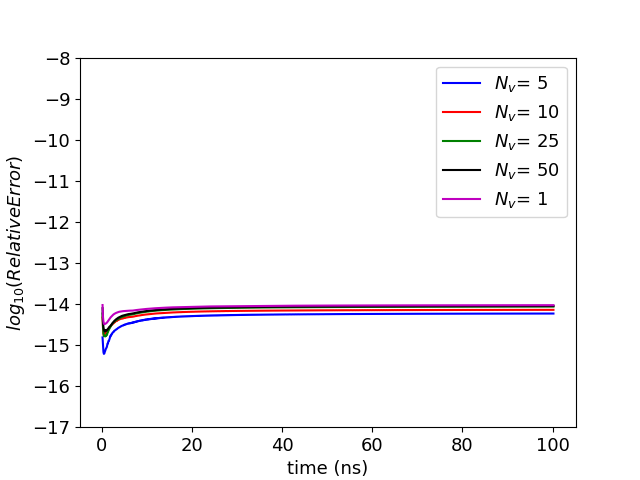}
\caption{ MFETI-DP2 }
\label{fig:dglF2}
\end{subfigure}
\caption{Error in Gauss' law for particle beam.}\label{fig:dgls}
\end{figure}

\begin{figure}
\centering
\begin{subfigure}{.85\linewidth}
\includegraphics[width=\linewidth]{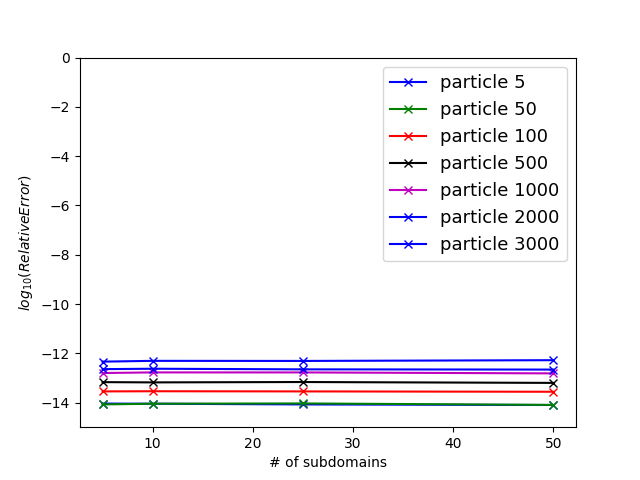}
\caption{ Difference in particle path for MFETI-DP1. }
\label{fig:pathF1}
\end{subfigure}
\begin{subfigure}{.85\linewidth}
\centering
\includegraphics[scale=.5]{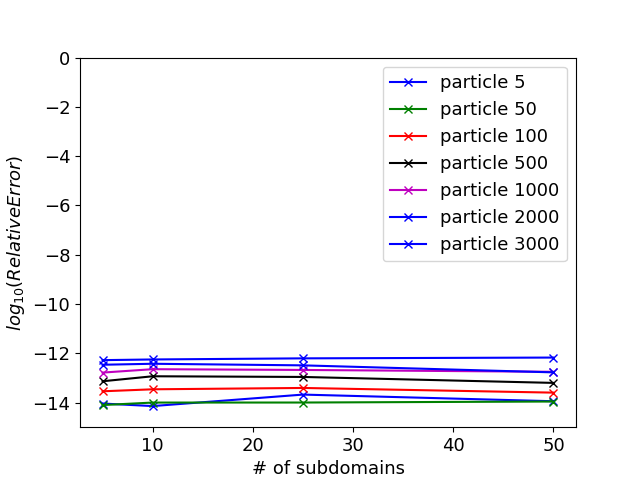}
\caption{ Difference in particle path for MFETI-DP2. }
\label{fig:pathF2}
\end{subfigure}
\caption{Error in tracked particle positions.}\label{fig:paths}
\end{figure}

The next two examples demonstrate MFETI-DP1 and MFETI-DP2 with particles.
In these experiments, the fields that are obtained through domain decomposition are used to push the particles, which are not subject to any decomposition scheme.
The first example is a particle beam traveling through a cylindrical PEC cavity.
This case is a standard test to show charge conservation as errors readily appear as striations in the particle distribution.
Furthermore, approximate solutions can be used to further verify the result.
The first test is done on a cavity that is 10 cm in length and 0.02 cm in radius disctretized with an average edge length of 5 mm resulting in roughly 21,000 unknowns.
The ten particles at each time step enter at the bottom face of the cavity with a radius of 8 mm.
The beam voltage and current are 500 V and 50 mA respectively, causing the an initial velocity $1.326\times 10^4$ km/s.
The geometry was subdivided in four configurations, the details of which are listed in Table \ref{tab:beam}.
The experiment was run for 1000 time steps at a time step size of 1 ns, which for the monolithic case took 1,342 seconds.
As shown in figure \ref{fig:dgls}, all the simulations maintained charge conservation over the course of the run.
There is no notable difference in the simulations, regardless of the number of subdivisions used.
Seven particles were tracked over the course of the run to track the particle trajectories with the domain decomposition.
Because the fields are virtually equivalent, the particle trajectories are consistent across the different geometries. Figure \ref{fig:paths} shows the error in the tracked particles' path with rep sect to the path in the monolithic solve.

\begin{figure}
\begin{subfigure}{.85\linewidth}
\centering
\includegraphics[scale=.5]{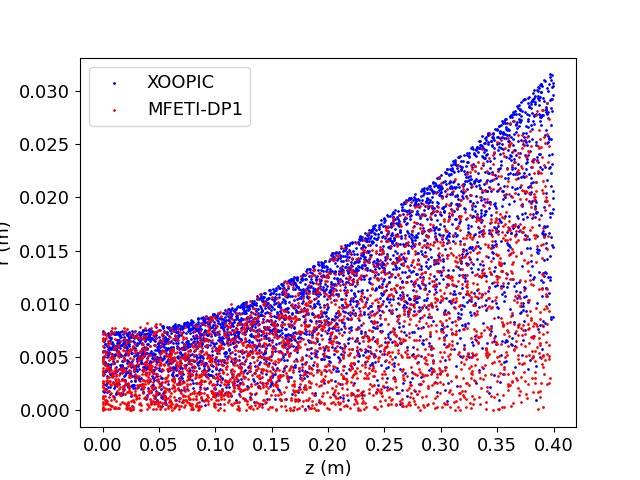}
\caption{ XOOPIC vs MFETI-DP1. }
\label{fig:beamF1}
\end{subfigure}
\begin{subfigure}{.85\linewidth}
\centering
\includegraphics[scale=.5]{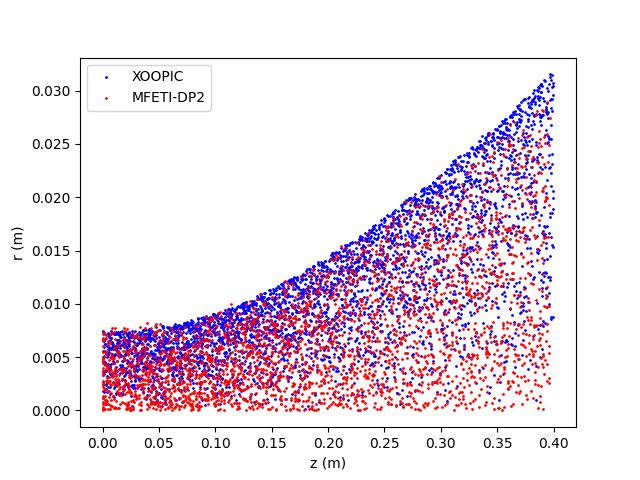}
\caption{ XOOPIC vs MFETI-DP2. }
\label{fig:beamF2}
\end{subfigure}
\caption{Particle beam expansion for enlarged PEC cavity.}\label{fig:bigBeam}
\end{figure}

The next result was a particle beam in a larger PEC cavity. 
In this case, the PEC cylinder was 40 cm in length and 10 cm in radius. 
The geometry has an average edge length of 7.38 mm and a total of 218,059 unknowns. 
The beam voltage and current are 3.2 kV and 50 mA respectively, resulting in an initial velocity of $3.35\times10^{4}$ km/s.
The experiment was run for both MFETI-DP1 and MFETI-DP2 using the iterative solver GMRES using a tolerance of $10^{-6}$. 
The particle distribution is compared to a well-validated axialsymmetric EM-FDTDPIC code, XOOPIC \cite{verboncoeur1995object} and shows good agreement in Figure \ref{fig:bigBeam}.

\subsection{Plasma Ball}
\begin{figure}
\begin{subfigure}{.85\linewidth}
\centering
\includegraphics[width=\linewidth]{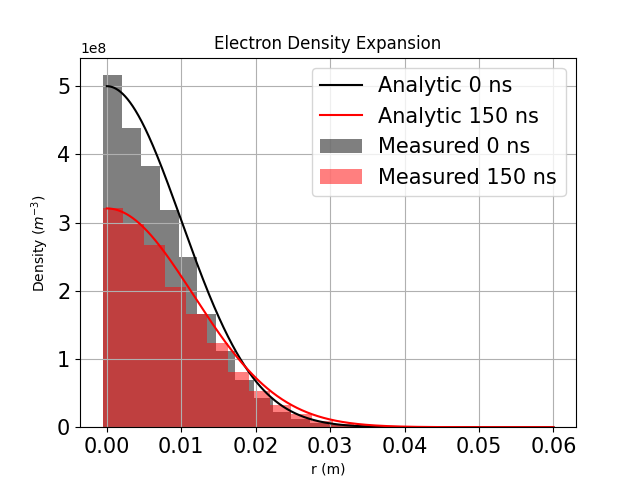}
\end{subfigure}
\begin{subfigure}{.85\linewidth}
\centering
\includegraphics[width=\linewidth]{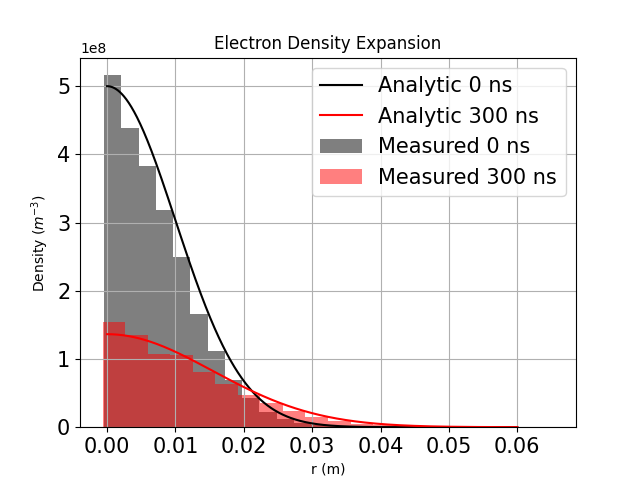}
\end{subfigure}
\begin{subfigure}{.85\linewidth}
\centering
\includegraphics[width=\linewidth]{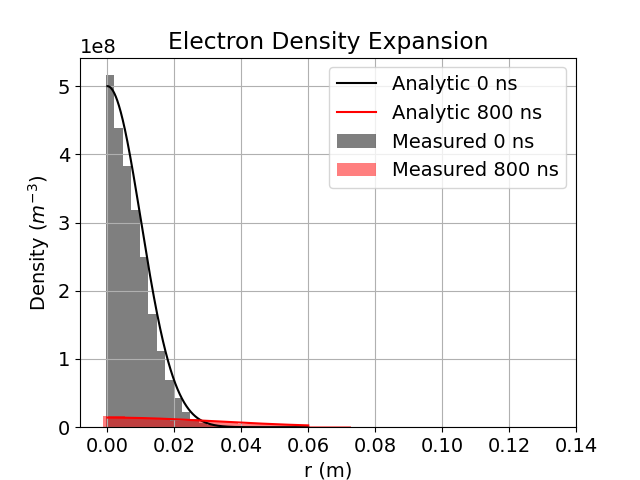}
\end{subfigure}
\caption{Adiabatic plasma ball expansion after 150, 300, and 800 ns.}\label{fig:pball}
\end{figure}

The last example presented is the case of an adiabatic expansion of a plasma ball using MFETI-DP2.
This example takes a considerable number or degrees of freedom to resolve the debye length and allow the particles to expand for an extended period of time. 
However, this can be made more tractable by using either domain decomposition schemes and unconditionally stable time marching to evolve the system with larger time steps than a similar EM-FDTDPIC formulation would allow.
Additionally, this experiment has analytic solutions \cite{kovalev2003analytic} that can be used for validation. 
A charge neutral, Gaussian distribution of 24000 particles is set at the center of a spherical region 24 cm in diameter.
The $Sr^+$ ions are at 1K and the electrons are at 100K with a density of $5\times10^8$ particles per cubic meter.
The outer boundary has an impedance boundary condition and the boundary is sufficiently far from the particles that none leave during the simulation.
The average edge length is 1 cm with a total of 681,214 unknowns.
Figure \ref{fig:pball} shows good agreement of the particle distribution with the analytic result at 150 ns, 300 ns, and 800 ns.

\section{Summary \label{sec:conclusions}}

In this work, we presented two formulations for domain decomposition using non-overlapping subdomains for the mixed finite element method. The methods were used to create a MFETI-PIC scheme which allows for computing the fields due to particle motion for larger problems.
For sufficiently large problems, this results in time savings without loss of accuracy.
All the results shown were computed on a single node; however, the method can be made parallel to take full advantage of high performance computing resources.
Doing so suggest creating a domain decomposition framework for particle mapping and pushing consistent with the field solver.

Future work will focus on creating a qausi-helmholtz decomposition for the FETI-framework. This ensures that the nullspace of the field solver does not corrupt the satisfaction of charge conservation. This is a key issue for using the time domain vector wave equation approach which has a nullspace that grows linearly in time. The combination of domain decomposition and qausi-Helmholtz decomposition will allow for solving even larger problems as the number of degrees of freedom is reduced to those associated with edges, rather than have both edges and faces as presented here.

\section{acknowledgments}
This work was sponsored by the US Air Force Research Laboratory under contracts FA8650-19-F-1747 and FA8650-20-C-1132.
This work was also supported by SMART Scholarship program. We thank the MSU Foundation for support through the Strategic Partnership Grant during early portion of this work. This work was also supported by the Department of Energy Computational Science Graduate Fellowship under grant DE-FG02-97ER25308. The authors would also like to thank the HPCC Facility, Michigan State University, East Lansing, MI, USA.

\bibliographystyle{IEEEtran}
\bibliography{ddm}

\end{document}